\documentclass[conference,10pt]{IEEEtran}
\usepackage{latexsym}
\usepackage{amsfonts}
\usepackage{amssymb}
\usepackage{amsbsy}
\usepackage{amsmath}
\usepackage{amsthm}
\usepackage{color}
\usepackage{dsfont}
\usepackage{pgf}
\usepackage{graphicx,lscape,rotating,subfig}
\usepackage{balance}
\usepackage{algorithm,algpseudocode}
\usepackage[noadjust]{cite}
\theoremstyle{remark}

\theoremstyle{definition}

\def\sinr{\mathsf{SINR}}

\def\({\left(}
\def\){\right)}

\def\[{\left[}
\def\]{\right]}

\def\b0{{\mathbf{0}}}

\def\cL{\mathcal{L}}

\newcommand{\Ebb}{\mathbb{E}}
\def\betad{\beta_\mathrm{d}}
\def\betau{\beta_\mathrm{u}}
\def\rhod{\rho_\mathrm{d}}
\def\rhou{\rho_\mathrm{u}}
\def\Td{T_\mathrm{d}}
\def\Tu{T_\mathrm{u}}

\def\tt{T_\mathrm{t}}

\def\lambdab{\lambda_\mathrm{b}}

\def\Pb{P_\mathrm{b}}
\def\Pm{P_\mathrm{m}}
\def\Sd{S_\mathrm{d}}
\def\Su{S_\mathrm{u}}

\newcommand\blfootnote[1]{%
  \begingroup
  \renewcommand\thefootnote{}\footnote{#1}%
  \addtocounter{footnote}{-1}%
  \endgroup
}

\title{Achieving Low Latency Two-Way Communication by Downlink and Uplink Decoupled Access}

\author{\IEEEauthorblockN{Dong Min Kim, Nuno K. Pratas and Petar Popovski}
\IEEEauthorblockA{Department of Electronic Systems, Aalborg University, Denmark\\
Email: \{dmk;nup;petarp\}@es.aau.dk} }

\begin{document}
\maketitle

\blfootnote{This work has been in part supported by the European Research Council (ERC
Consolidator Grant Nr. 648382 WILLOW) within the Horizon 2020 Program. Part of this work
has been performed in the framework of the Horizon 2020 project ONE5G (ICT - 760809)
receiving funds from the European Union. The authors would like to acknowledge the
contributions of their colleagues in the project, although the views expressed in this
contribution are those of the authors and do not necessarily represent the project.}

\begin{abstract}
In many scenarios, low latency wireless communication assumes two-way connection,
such that the node that receives information can swiftly send acknowledgment or
other response. In this paper, we address the problem of low latency two-way
communication and address it through proposal of a base station (BS) cooperation
scheme. The scheme is based on downlink (DL) and uplink (UL) decoupled access
(DUDA). To the best of our knowledge, this is the first time that the idea of
decoupled access is used to reduce latency.  We derive the analytical expression
for the average latency and verify that the latency expression is valid with
outage probability based on stochastic geometry analysis. Both analytical and
simulation results show that, with DUDA, the latency can be reduced by
approximately 30-60\% compared to the traditional coupled access.
\end{abstract}


%\begin{proposition}\label{prop1} Text text ... \end{proposition}
%\begin{proof} Text text ... \end{proof}
%\begin{lemma} Text text ... \end{lemma}
%\begin{lemma} Text text ... \end{lemma}
%\begin{thm} Text text ... \end{thm}
%\begin{thma} Text text ... \end{thma}
%\begin{rem} Text text ... \end{rem}
%\begin{definition} Text text ... \end{definition}

\section{Introduction}

One of the most promising use cases of the emerging 5G wireless systems is the one
of reliable low latency communications \cite{imtvision,tr38802}, such as
vehicle-to-vehicle (V2V) and vehicle-to-anything (V2X) connections
\cite{ashraf2017towards}, ultra-reliable connections in industrial environments
\cite{schulz2017latency}. A reliable communication link is established through a
two-way communication protocol, where the receiver always acknowledges the reception
of the transmitter's packet. This brings forward the need to have two-way
communication links where each device can quickly switch between transmission and
reception. The obvious way to achieve low latency is to use full duplex transceivers
through the use of Frequency Division Duplex (FDD) systems. Contrary to this, many
ongoing efforts are currently favoring Time Division Duplex (TDD) operation in order
to reduce transceiver cost, increase spectrum usage efficiency, take advantage of
channel reciprocity and to be capable to adapt to time-varying uplink/downlink
traffic asymmetries \cite{shen2012dynamic}. However, the frame-based structure of
TDD is not aligned with the requirement for latency reduction due to the long time
it takes to switch between uplink and downlink. Based on these observations, we
conclude that there is a need to design low latency two-way communication solutions
that can work with terminals (devices) that operate in TDD.

\begin{figure}[tb]
\centering
\subfloat[]{
\includegraphics[width=\linewidth]{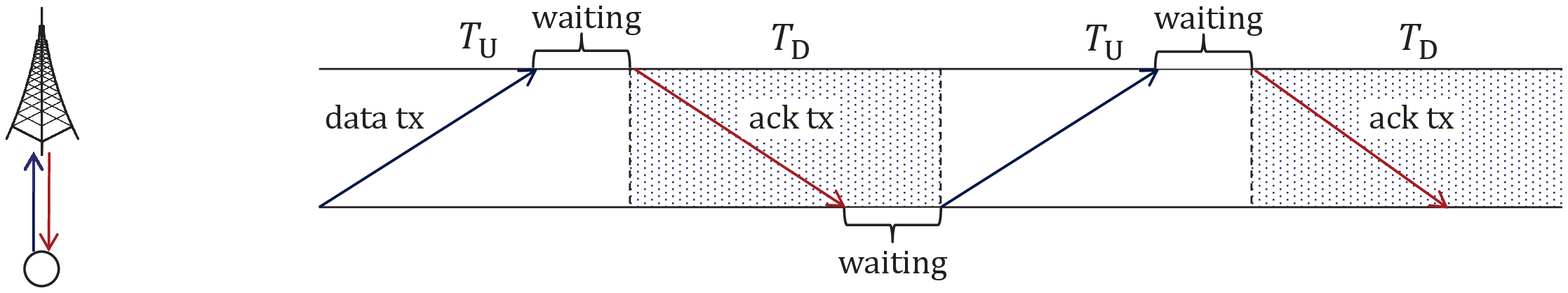}
\label{F:intro_example}}
\\
\subfloat[]{
\includegraphics[width=\linewidth]{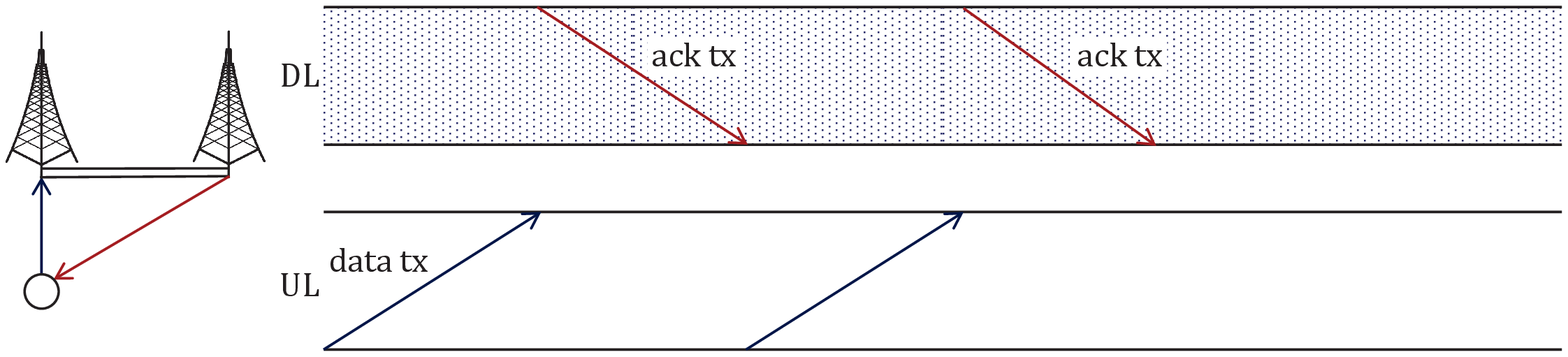}
\label{F:proposed_example}}
\caption{Two way traffic initiated in the uplink: (a) $T_{\rm{U}}$ denotes uplink frame duration and $T_{\rm{D}}$ donwlink frame duration; (b) Proposed scheme with two half-duplex base-stations, where one takes the role of uplink and the other of downlink.}
\label{F:TDD_example}
\end{figure}

In TDD cellular systems, the minimization of the latency experienced by a two-way
communication transaction is constrained by the time period (frame duration) that
the half-duplex base station stays in the downlink and uplink directions as shown in
Fig.~\ref{F:intro_example}. The direct approach is to decrease the time period
before shifting from uplink (downlink) to downlink (uplink) direction, in order to
facilitate fast interaction between the communicating devices/nodes
\cite{mahmood2016radio,pocovi2016impact}. This fast switching between uplink and
downlink comes at the cost of more complex transceivers both at the device and base
station, due to the need to perform faster and more frequent channel estimation, as
well as additional signaling overhead, due to the resource assignment.

This paper proposes a radically different method for enabling low-latency two-way
communication in frame-based TDD cellular systems. We propose to use two half-duplex
base stations instead of one, where the first base station takes the downlink
direction and the second the uplink direction (or vice versa). As a motivation
example on how this approach can satisfy the individual low-latency requirements,
consider the example in Fig.~\ref{F:intro_example} with two devices where each
should receive a reply from their original transmission within 2 time-slots. The TDD
structure of the baseline scheme in Fig.~\ref{F:intro_example} is not capable of
avoiding additional waiting times; while, as shown in Fig.~\ref{F:proposed_example},
the proposed scheme can support the desired latency, as it allows each device to
switch between uplink and downlink directions without requiring the network
infrastructure to do so. This of course assumes that the processing time within the
device to switch between UL and DL directions is lower than the residual time
remaining in the traditional cellular TDD to switch between UL and DL (and
vice-versa).

The proposed scheme can be applied both in a traditional and in a cloud radio access
network (C-RAN) architecture. In the traditional architecture, the coordination
between the half-duplex base stations can be accomplished through the X2 interface;
while in a C-RAN architecture this coordination is implicit at the C-RAN's base-band
unit (BBU). The interference resulting from the downlink to the uplink half-duplex
base stations can be mitigated by: (i) Taking advantage of spatial pre-computing,
beamforming and full-dimension MIMO to steer the downlink interfering beam from the
uplink base station; and/or (ii) take advantage of the X2 interface to exchange the
necessary information to cancel any residual wireless interference, with the goal of
improving the reliability of the uplink reception in a low latency traffic setting.

The proposed scheme builds upon already on-going efforts in 3GPP such as device
multi-connectivity, decoupled uplink and downlink access \cite{boccardi2016why},
cooperation between base stations and dynamic TDD. Furthermore, it enables the
network to continue its evolution towards a device centric architecture and enable
the joint design and scheduling of low latency two-way communications
\cite{popovski2015spin}.

To the best of our knowledge, this is the first study to apply decoupled access in order to decrease latency.
We propose a system design, which enables low latency two-way interactive
      communications in a TDD regime, without incurring the latency penalties
      associated with the TDD UL and DL cycling periods. This is of relevance
      since TDD operation allows to reduce transceiver cost, increase spectrum
      usage efficiency and adapt to time-varying Uplink/Downlink (UL/DL) traffic
      asymmetries \cite{shen2012dynamic}.
We quantify the proposed scheme and we show that the proposed scheme is
      always able to achieve lower two-way communication latency than the baseline
      scheme.

The paper is organized as follows. We describe the system model and the proposed
scheme in Section~II and Section~III, respectively. The latency and reliability
analysis of proposed scheme is given in Section~IV, and its numerical results are
presented in Section~V. The paper is concluded in Section~VI.

\section{System Model}

In this section, we explain the system model to evaluate the latency performance of
two-way traffic in the cellular networks. some basic assumptions are provided in the
following subsections.

\begin{figure}[tb]
\centering
\subfloat[]{
\includegraphics[height=0.3\linewidth]{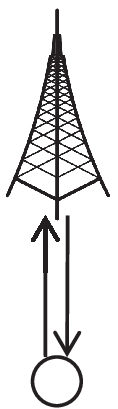}
\label{F:concept_DUCA}}
\subfloat[]{
\includegraphics[height=0.3\linewidth]{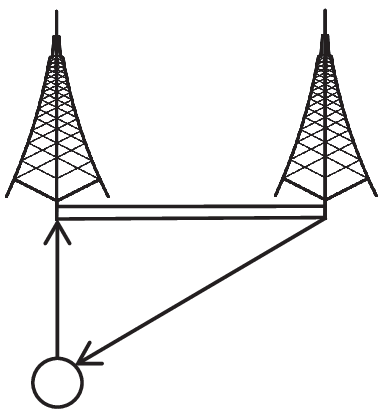}
\label{F:concept_DUDA}}
\caption{(a) Baseline scheme; (b) Proposed scheme.}
\label{F:concept}
\end{figure}

\subsection{Network and Channel Model}

In a cellular network, the base stations (BSs) are randomly distributed with density
$\lambdab$, resulting in a homogeneous Poisson point process (PPP)
\cite{Chiu2013stochastic}. A pair of BSs are interconnected via a wired connection
(double solid line in Fig.~\ref{F:concept_DUDA}). BSs can serve cross directional
traffic, one BS will operate in DL (DL-BS) and the other in UL (UL-BS), or
vice-versa. The UL-BS can use side information sent from the DL-BS through the wired
backhaul. We assume that the closer BS will operate in UL (UL-BS) and the other in
DL (DL-BS).

The user equipments (UEs) associate with their nearest BSs first. A single spectrum
with unit bandwidth and Rayleigh fading channel with unit mean power is assumed. All
transmitted signals experience path loss as follows: $\ell(r)=r^{-\alpha}$, with
path loss exponent $\alpha$. The default transmit power of uplink UEs is $\Pm$. BSs
transmit with constant power $\Pb$.

\subsection{Transmission Success and Retransmission Model}

A signal-to-interference-plus-noise ratio (SINR) requirement for uplink and downlink
is denoted by $\betau$ and $\betad$, respectively. The transmission is successful if
the received SINR is greater than or equal to the target threshold value:
\begin{align}
\rhod &\triangleq \Pr\[\sinr_\mathrm{d} > \beta_\mathrm{d}\], \label{E:rhod} \\
\rhou &\triangleq \Pr\[\sinr_\mathrm{u} > \beta_\mathrm{u}\]. \label{E:rhou}
\end{align}
Even if the transmission is successful, if the acknowledgement is not received, a
retransmission is attempted. Retransmission stops when the maximum number of
retransmissions is reached or an acknowledgement arrives. With $n$ transmission
opportunities, the transmission success is described as follows:
\begin{align}\label{E:psn}
p_\mathrm{s}^{(n)} \triangleq \sum_{i=0}^{n-1}\(1-\rhod\rhou\)^i \rhod\rhou = 1 - \(1-\rhod\rhou\)^n.
\end{align}

\subsection{Traffic and Association Model}

In this paper, we consider a situation where data is transmitted on the UL and its
acknowledgment (ACK) is transmitted on the DL. If the UL transmission fails, the BS
does not transmit an ACK, so the UE waits for the ACK for a certain period of time
and retransmits the data. The same operation is performed when the BS transmits an
ACK, but this is not decoded successfully at the UE.

In general the transmission power output of the UE is weaker than that of the BS.
So, it is efficient to allocate a better channel to the UL transmission. Let us set
up that the UE receives the UL transmission from the far BS and transmits the UL
data to the nearer BS among two cooperating BSs.

\section{Proposed Scheme}

The basic principle of our proposal is the introduction of two Half-Duplex Base
Stations (HDBS), the first operating in the downlink and the second on the uplink
(or vice-versa), as depicted in Fig.~\ref{F:concept_DUDA}. A half-duplex mobile
device that connects to the infrastructure, associates simultaneously with both
HDBS, such that at one time it receives from one of the HDBSs and at another time it
transmits to the other HDBS.

This scheme will permit to reduce the latency of two-way communication transactions
beyond what is possible using a traditional TDD system. It should be noted that this
is in line with the recent trends of decoupled uplink/downlink access
\cite{boccardi2016why}, device multi-connectivity, cooperation between base stations
and dynamic TDD; however, the instance and the benefit presented in this paper have
not been observed so far.

In TDD systems, the proposed scheme, where the Half-Duplex Base Stations (HDBSs) are
connected via an X2 interface (Fig.~\ref{F:concept_DUDA}), can achieve lower
latencies than the single TDD baseline (Fig.~\ref{F:concept_DUCA}). The baseline
scheme is composed by a half-duplex device and a HDBS as shown in
Fig.~\ref{F:concept_DUCA}. The simplest realization of the proposed scheme is
depicted in Fig.~\ref{F:concept_DUDA}. Each transceiver on the picture, both at the
UE and at the infrastructure, is half-duplex and operates in a TDD mode.

\section{Latency and Reliability Analysis}

In the following we show that the latency analysis of the baseline and proposed
scheme.

\subsection{Latency Model}

\subsubsection{Protocol delay}

\begin{figure}[tb]
    \centering
        \includegraphics[width=0.4\linewidth]{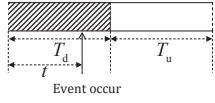}
	\caption{Illustration of time slot composition.}
\label{F:timeslot}
\end{figure}

For TDD system, the protocol delay is the waiting time until the transmission slot
arrives. The time is discretized into downlink and uplink time slot with durations
$\Td$ and $\Tu$, respectively.

We assume that the packet is generated randomly and uniformly over the time at the
UE. If a packet with size\footnote{The size of the packet is measured by the amount
time it occupies.} $\Su \(0<\Su\leq\Tu\)$ is generated after $t \(0\leq
t\leq\Td+\Tu\)$ time from the start of the DL slot, the protocol latency
$L_\mathrm{p}$ can be described as:
\begin{align}\label{E:Lp}
L_\mathrm{p} &= \(\Td-t\)\cdot\Pr\(t \leq \Td\) + 0\cdot\Pr\(\Td < t \leq \Td+\Tu-\Su\) \nonumber \\
&+\(\Td+\Tu-t+\Td\)\cdot\Pr\(\Td+\Tu-\Su < t \leq \Td+\Tu\) \nonumber \\
&=\(\Td-t\)\frac{\Td}{\Td+\Tu} +\(2\Td+\Tu-t\)\frac{\Su}{\Td+\Tu}.
\end{align}

The number of transmissions until success can be modeled by geometric distribution.
Then the retransmission delay $L_\mathrm{r}$ can be described as:
\begin{align}\label{E:Lr}
L_\mathrm{r} &= \(\Td + \Tu\)\cdot\(\frac{1}{\rhou\rhod}-1\).
\end{align}
Even without protocol delay and retransmission delay, there is a fundamental delay
$L_\mathrm{f}$ caused from the transmission delay and reception delay. If a size of
the acknowledgment in downlink is $\Sd \(0<\Sd\leq\Td\)$, the fundamental delay can
be described as:
\begin{align}\label{E:Lf}
L_\mathrm{f} = \Tu + \Sd.
\end{align}

Then, the total expected latency of TDD scheme (or downlink and uplink coupled
access (DUCA)) $L_\mathrm{DUCA}$ is:
\begin{align}\label{E:Lduca}
L_\mathrm{DUCA} &= \Ebb\[L_\mathrm{p} + L_\mathrm{r} + L_\mathrm{f}\] \nonumber \\
&=\frac{\Td^2+\(2\Td+\Tu\)\Su}{\Td + \Tu} - \frac{\Td + \Su}{2} \nonumber \\
&\quad+ \(\Td + \Tu\)\(\frac{1}{\rhou\rhod}-1\) + \Tu + \Sd.
\end{align}

For the proposed scheme, because the user can transmit its packet whenever, there is
no protocol delay caused by fixed time division duplex. Then, only retransmission
delay and fundamental delay remain. But there is a waiting time $W$ for ACK. It is
fair to set this time period as same as $\Td$:
\begin{align}\label{E:Lduda}
L_\mathrm{DUDA} &= \(\Su + W\)\(\frac{1}{\rhou\rhod}-1\) + \Su + \Sd \nonumber \\
&= \(\Su + \Td\)\(\frac{1}{\rhou\rhod}-1\) + \Su + \Sd.
\end{align}

Let us compute $L_\mathrm{DUCA} - L_\mathrm{DUDA}$. With assuming $\Td=\Tu$ to be
fair.
\begin{align}\label{E:LducaminusLduda}
L_\mathrm{DUCA} - L_\mathrm{DUDA} &= \frac{\Tu-\Su}{\rhou\rhod} + \Su > 0.
\end{align}
So, the proposed scheme is always able to achieve lower two-way communication
latency than the baseline scheme.

\subsection{Reliability analysis}

In this subsection, we investigate the reliability performance by deriving
analytical expressions for the transmission success probabilities in UL and DL using
stochastic geometry. For the analytical tractability, we assume that all BSs in the
network can schedule one UE either UL (UL-UE) or DL (DL-UE) in each Voronoi cell for
the TDD case. For the proposed scheme, a pair of cooperating BSs will serve only one
UE at a time. We further assume that the spatial distribution of UEs follows another
independent PPP with the density $\lambdab$.

\subsubsection{UL Success Probability in DUDA Network}

The success probability of the transmission of a typical UL user $\mathcal{U}$ at a
typical BS $B\(\mathcal{U}\)$ in DUDA $p_\mathrm{U}^\mathrm{DUDA}$ can be expressed
as follows:
\begin{align}\label{E:sucprobULDUDA}
p_\mathrm{U}^\mathrm{DUDA} &= \Ebb\left[ \Pr \left[ \frac{g_{\mathcal{U},B\(\mathcal{U}\)}r^{-\alpha}\Pm}{I_{B\(\mathcal{U}\)}^\psi+I_{B\(\mathcal{U}\)}^\varphi} \geq \betau \right] \right] \\
    &= \Ebb\left[ \Pr \left[ g_{\mathcal{U},B\(\mathcal{U}\)} \geq \frac{\betau r^\alpha}{\Pm} \(  I_{B\(\mathcal{U}\)}^\psi+I_{B\(\mathcal{U}\)}^\varphi \) \right] \right], \nonumber
\end{align}
where $g_{i,j}$ denotes the gain at node $j$ of the channel from node $i$, and
$I_{i}^\psi$ and $I_{i}^\varphi$ denote the aggregate interference at node $i$ from
DL-BSs and UL-UEs, respectively. The UL transmission distance between the typical
user and the typical BS is denoted by $r$, and $r$ is assumed as random variable
with pdf as $f\(r\)=2\pi \lambdab r \exp\(-\pi \lambdab r^2\)$
\cite{andrews2011tractable}. Due to the Rayleigh faded channel,
$g_{\mathcal{U},B\(\mathcal{U}\)}$ is an exponential random variable with unit mean,
then \eqref{E:sucprobULDUDA} can be expressed as:
\begin{align}\label{E:sucprobULDUDA2}
p_\mathrm{U}^{\mathrm{DUDA}}
    &=\Ebb_r\left[
            \Ebb_{I_{B\(\mathcal{U}\)}^\psi,I_{B\(\mathcal{U}\)}^\varphi}\left[
                \exp\( -s \( I_{B\(\mathcal{U}\)}^\psi + I_{B\(\mathcal{U}\)}^\varphi \) \)\right]
          \right] \nonumber \\
    &=\Ebb_r\left[
            \Ebb_{I_{B\(\mathcal{U}\)}^\psi}\left[
                e^{-sI_{B\(\mathcal{U}\)}^\psi}\right]
            \Ebb_{I_{B\(\mathcal{U}\)}^\varphi}\left[
                e^{-sI_{B\(\mathcal{U}\)}^\varphi}\right]
          \right] \nonumber \\
    &=\int_0^\infty   \cL_\mathrm{U}^\psi \left( s \right) \cL_\mathrm{U}^\varphi \left( s \right) 2\pi {\lambdab}r\exp \left( { - \pi {\lambdab}{r^2}} \right)dr,
\end{align}
where $s = \frac{\betau r^\alpha}{\Pm}$, and $\cL_\mathrm{U}^\psi \left( s \right)$
and $\cL_\mathrm{U}^\varphi \left( s \right)$ are the Laplace functionals of the
interference from DL-BSs at the typical BS and the interference from UL-UEs at the
typical BS, respectively. The interference from DL-BS is coming from outside of the
pair. Assuming independence of the channels from different interfering DL-BSs and
independence of the distances from different interfering DL-BSs, and using moment
generating function of exponential distribution, $\cL_\mathrm{U}^\psi \left( s
\right)$ can be expressed as:
\begin{align}
\label{E:ltpsiul1}
\cL&_\mathrm{U}^\psi \left( s \right)
    = \Ebb_{I_{B\(\mathcal{U}\)}^\psi}\left[ { \exp\(-\frac{{\betau}{r^\alpha}}{\Pm} I_{B\(\mathcal{U}\)}^\psi \) } \right] \nonumber \\
    &= \Ebb_{g_{i,B\(\mathcal{U}\)},r_{i,B\(\mathcal{U}\)}}\left[ { \exp\(-\frac{{\betau}{r^\alpha}}{\Pm} \sum\limits_{i \in \Phi^\psi}{g_{i,B\(\mathcal{U}\)}} r_{i,B\left( \mathcal{U} \right)}^{ - \alpha }\Pb \) } \right] \nonumber \\
    &= \Ebb_{r_{i,B\(\mathcal{U}\)}}\!\left[ {\prod\limits_{i \in \Phi^\psi} \Ebb_{g_{i,B\(\mathcal{U}\)}}\! \left[\exp\!\(\!-\frac{\Pb}{\Pm} {g_{i,B\(\mathcal{U}\)}} {\betau}{r^\alpha }r_{i,B\left( \mathcal{U} \right)}^{ - \alpha }\)\right] } \right] \nonumber \\
    &= \Ebb_{r_{i,B\(\mathcal{U}\)}}\left[ {\prod\limits_{i \in \Phi^\psi} {\frac{1}{{1 + \left( {{\Pb}/{\Pm}} \right){\betau}{r^\alpha }r_{i,B\left( \mathcal{U} \right)}^{ - \alpha }}}} } \right],
\end{align}
where $r_{i,B\(\mathcal{U}\)}$ denotes the distance from $i$th BS in the interfering
BS set $\Phi^\psi$ to the typical BS $B\(\mathcal{U}\)$. To model the density of
interfering nodes, we define $\delta$ as the ratio of DL traffic of the entire
network. The traffic asymmetry can be modeled by adjusting $\delta$. The range of
$\delta$ is $(0, 1)$. We assume that the value of $\delta$ is not changed over the
observation period even though the real transmitting nodes are varying. The average
node density of DL-BSs is $\delta\lambdab$. Because a pair of cooperating BSs will
serve one user, the higher achievable node density of the DL-BS is $0.5\lambdab$. We
assume that the set $\Phi^\psi$ follows a PPP with density $\lambdab^\psi$, where
$\lambdab^\psi=0.5\delta\lambdab$. Assuming the paired BS is the nearest DL-BS of
the UL-BS, the distance to the nearest \emph{interfering} BS is the distance to the
second nearest DL-BS. The second nearest distance distribution is given as
\cite{moltchanov2012distance}:
\begin{align}\label{E:2ndnearest}
f\(d\)=2\(\pi\lambda\)^2 d^3 \exp\(-\pi\lambda d^2\).
\end{align}
Using \eqref{E:2ndnearest}, $\lambdab^\psi$, and probability generating functional
(PGFL) of PPP \cite{Chiu2013stochastic}, \eqref{E:ltpsiul1} can be expressed as:
\begin{align}
\label{E:ltpsiul2}
\cL_\mathrm{U}^\psi \left( s \right) &=\! \int_0^\infty \!\!
    \exp \left( { - 2\pi \lambdab^\psi
            \int_t^\infty \! {\frac{{\left( {{\Pb}/{\Pm}} \right){\betau}{r^\alpha }{x^{ - \alpha }}}}{{1 + \left( {{\Pb}/{\Pm}} \right){\betau}{r^\alpha }{x^{ - \alpha }}}}xdx} }
         \right)\cdot \nonumber \\
&\qquad\qquad 2 \(\pi \lambdab\)^2 t^3 \exp \left( { - \pi \lambdab {t^2}} \right)dt,
\end{align}
where $t$ is the distance to the nearest interfering DL-BS (second nearest BS). For
brevity, $r_{i,B\(\mathcal{U}\)}$ is changed as $x$.

In a similar way, we can obtain $\cL_\mathrm{U}^\varphi \left( s \right)$. The
interfering MS set will be denoted $\Phi^\varphi$ and it is assumed as PPP with
density $\lambdab^\varphi$.
The average node density of UL-MSs is
$0.5\(1-\delta\)\lambdab$. So, $\lambdab^\varphi$ is equal to
$0.5\(1-\delta\)\lambdab$. Then $\cL_\mathrm{U}^\varphi \left( s \right)$ can be
expressed as:
\begin{align}
\label{E:ltvarphiul}
&\cL_\mathrm{U}^{\varphi}(s) = \exp \left( { - 2\pi \lambdab^\varphi \int_r^\infty  {\frac{{{\betau}{r^\alpha }{y^{ - \alpha }}}}{{1 + {\betau}{r^\alpha }{y^{ - \alpha }}}}ydy} } \right),
\end{align}
where $y$ denotes the distance from interfering UL-MSs to the typical BS. It is
assumed that the interfering UL-MSs are located at a distance larger than $r$. Using
\eqref{E:sucprobULDUDA}, \eqref{E:ltpsiul2}, and \eqref{E:ltvarphiul}, we can
evaluate the UL transmission success probability for DUDA.

\subsection{DL Success Probability in DUDA Network}

Following the same lines as $p_\mathrm{U}^\mathrm{DUDA}$, we can obtain the
analytical expression for the DL transmission success probability as follows:
\begin{align}\label{E:sucprobDLDUDA}
p_\mathrm{D}^\mathrm{DUDA} \approx \int_0^\infty  {\cL_\mathrm{D}^\psi \left( s \right)\cL_\mathrm{D}^\varphi \left( s \right)2\pi {\lambdab}^2 r^3\exp \left( { - \pi {\lambdab}{r^2}} \right)dr},
\end{align}
where $s = \frac{\betad r^\alpha}{\Pb}$ and the Laplace functionals of the
interference from BSs $\cL_\mathrm{D}^{\psi}\(s\)$ and MSs
$\cL_\mathrm{D}^{\varphi}\(s\)$ are
\begin{align}
\label{E:ltpsidl}
\cL_\mathrm{D}^{\psi}\(s\) &= \exp \left( { - 2\pi {\lambdab^\psi}\int_r^\infty  {\left( {\frac{{{\betad}{r^\alpha }{x^{ - \alpha }}}}{{1 + {\betad}{r^\alpha }{x^{ - \alpha }}}}} \right)xdx} } \right), \nonumber \\
\end{align}
\begin{align}
\label{E:ltvarphidl}
\cL_\mathrm{D}^{\varphi}\(s\) = \exp \left( { - 2\pi {\lambdab^\varphi}\int_0^\infty  {\frac{{\left( {{\Pm}/{\Pb}} \right){\betad}{r^\alpha }{y^{ - \alpha }}}}{{1 + \left( {{\Pm}/{\Pb}} \right){\betad}{r^\alpha }{y^{ - \alpha }}}}ydy} } \right).
\end{align}
The distance from the interfering DL-BSs to the typical DL-MS $x$ cannot be closer
than $r$ as shown in \cite{andrews2011tractable}. We further apply this distance
restriction to the distance from the interfering UL-MSs to the typical DL-MS $y$ to
approximate.

\section{Numerical Results}

We quantify the latency performance of DUDA, where the analytical results of
\eqref{E:sucprobULDUDA} and \eqref{E:sucprobDLDUDA} are compared with numerical
simulations. All the simulation results are obtained by performing Monte Carlo
simulation with 10000 iterations. The common parameters used are shown in
Table~\ref{T:SimulationParameters}.

\begin{table}[b]
	\caption{Simulation parameters.}
	\centering
			\begin{tabular}{ c l c }
		  	\hline
  			Parameter & Description & Simulation Setting\\
				\hline
				$S$ & Size of observation window & $150$ m \\
				$\lambda_\mathrm{b}$ & BS density & $0.005$ ${\text{BS}}/{\text{m}^2}$ \\
				$\delta$ & Traffic asymmetry ratio & $0.5$  \\
				$\sigma^2$ & Noise power at MS and BS & $-174$ dBm \\
				$\alpha$ & Path loss exponent & $4$ \\
				$\betad$ & DL SINR threshold (ACK) & -5 dB \\
				$\betau$ & UL SINR threshold (Data) & 0 dB \\
				$\Pb$ & BS transmission power & $40$ dBm \\
				$\Pm$ & MS transmission power & $20$ dBm \\
				$W$ & System bandwidth & $1$ Hz \\
				$N$ & Simulation iterations & 10000 \\
				\hline
			\end{tabular}
	\label{T:SimulationParameters}
\end{table}

The analysis from \eqref{E:LducaminusLduda} implied the proposed method has superior
performance in terms of latency, measured in time slots.  This is further verified
through a simulation, conducted in a two-dimensional spatial region, following the
principles for wireless network evaluation through stochastic geometry models.
The simulation is built as follows:
\begin{enumerate}
  \item Deploy HDBSs according to Poisson point process (PPP) with intensity
      $\lambda_\mathrm{b}$. The typical node (DL: UE, UL: BS) is located at the
      origin.
  \item Construct Voronoi cell.
  \item Starting from one HDBS randomly, search nearest HDBS among HDBSs who are
      sharing the same edges.
  \item Make a pair if the HDBS is unpaired.
  \item For the interfering BS pairs, randomly generate transmission direction
      according to $\delta$ and deploy active UE in one of cells of cooperating
      BSs.
  \item Repeat the procedure 3)-5) for the other HDBS randomly.
  \item Calculate SINR and until two-way transmission succeeds.
  \item Record latency and return to step 1) and repeat 10000 times.
\end{enumerate}

\begin{figure}[tb]
	\centering
		\includegraphics[width=\linewidth]{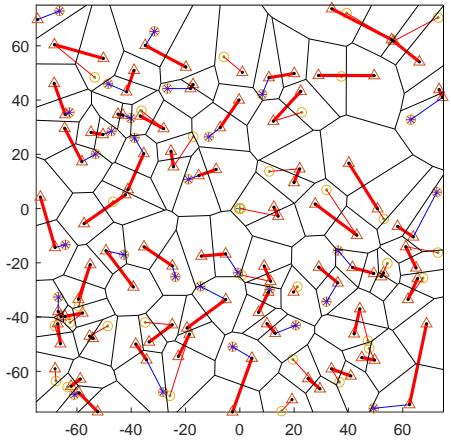}
	\caption{A snapshot of proposed scheme in two dimensional network.}
	\label{F:topology_duda}
\end{figure}

Only one of the cooperating BSs has an active UE to make two BSs serving one UE. A
snapshot of the deployment is as shown in Fig.~\ref{F:topology_duda}.

\begin{figure}[tb]
	\centering
		\includegraphics[width=\linewidth]{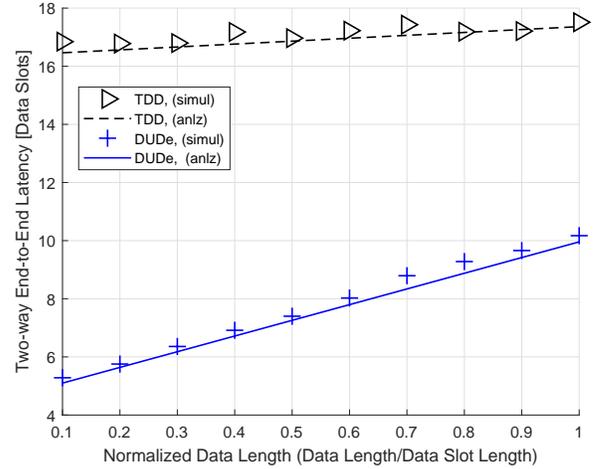}
	\caption{Two-way latency as a function of normalized data length.}
	\label{F:latency_datalength_sg}
\end{figure}

\begin{figure}[tb]
	\centering
		\includegraphics[width=\linewidth]{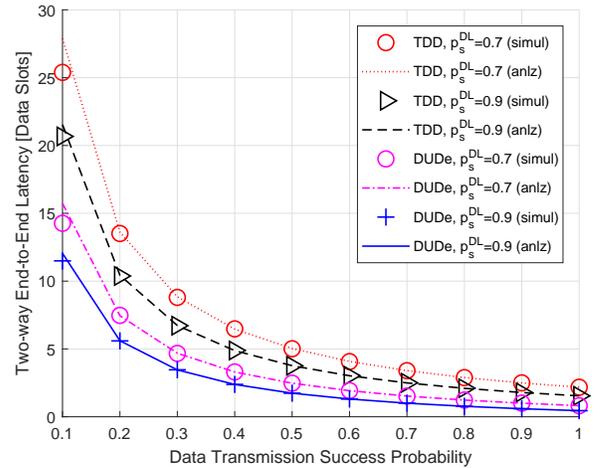}
	\caption{Two-way latency as a function of data transmission success probability.}
	\label{F:latency_sucprob}
\end{figure}

In Fig.~\ref{F:latency_datalength_sg}, we validate the our latency analysis with the
normalized data length. Because the data slot length is one, the normalized data
length has the value in $(0, 1)$ range. As the normalized data length becomes
longer, the transmission time increases. Therefore, it is natural that transmission
delay increases in both DUCA and DUDA in proportion to the normalized data length.
However, in the case of DUDA, it shows shorter latency of about 30--60\% compared to
DUCA. This is because there is no time slot switching time in two-way transmission.
In the case of DUCA the UE always communicates with the nearest BS, which means that
UE and BS are the optimal combination. However, the DUDA communicates with the
nearest BS in UL, but not in DL. As we can see from
Fig.~\ref{F:latency_datalength_sg}, this seems to
have little effect.

As we have seen the above, DUDA does not always utilize the topology that produces
optimal communication. That is, in some cases the probability of successful
transmission may be very low. The impact of the transmission success probability is shown on Fig.~\ref{F:latency_sucprob}. The two-way latency
is expressed as a function of the transmission success probability. The latency is
inversely proportional to the success probability of the data transmission, since
additional retransmissions are required if the lower is the success transmissions of
the data transmission. Both DUCA and DUDA show similar trends. However, in all cases
where the simulation was performed, the DUDA exhibits a lower latency than the DUCA.
The difference in performance gap between the two is larger when the transmission
success probability is low. This matches the expected behaviour as it was shown in
the derivation of \eqref{E:LducaminusLduda}.

\section{Concluding Remarks}

In this paper, the latency performance of the downlink and uplink decoupled access
(DUDA) was investigated. We verified that the latency expression is valid with
outage probability based on stochastic geometry analysis. Both analytical and
simulation results showed that the latency performance is improved with DUDA. The
proposed scheme can be applied to improve the current ongoing LTE-TDD and its
evolution towards 5G.

% Generated by IEEEtran.bst, version: 1.14 (2015/08/26)

%\bibliographystyle{IEEEtran}  % appearance order
%\bibliography{latency_twoway}

\end{document}